\documentclass[preprint,aps,nofootinbib]{revtex4-1}
\pdfoutput=1

\usepackage{amsmath}  
\usepackage{amsbsy,amssymb,amsfonts}
\usepackage{graphicx}
\usepackage{color}
\usepackage{epsf}
\usepackage[utf8x]{inputenc}
\usepackage{hyperref}
\usepackage{multirow}
\def\cm{$\rm cm^{-1}$}

\def\bravert{\egroup\,\vrule\,\bgroup}

{\catcode`\|=\active
  \gdef\Twoint#1{\left(\mathcode`\|"8000\let|\bravert {#1}\right)}}
{\catcode`\|=\active
  \gdef\Braket#1{\left<\mathcode`\|"8000\let|\bravert {#1}\right>}}

\newcommand{\beq}{\begin{equation}}
\newcommand{\eeq}{\end{equation}}
\newcommand{\beqa}{\begin{eqnarray}}
\newcommand{\eeqa}{\end{eqnarray}}
\newcommand{\bea}{\begin{array}}
\newcommand{\eea}{\end{array}}

\newcommand{\bef}{\begin{figure}}
\newcommand{\ef}{\end{figure}}
\newcommand{\bc}{\begin{center}}
\newcommand{\ec}{\end{center}}
\newcommand{\bt}{\begin{table}}
\newcommand{\et}{\end{table}}
\newcommand{\btb}{\begin{tabular}}
\newcommand{\etb}{\end{tabular}}

\def\rvac{\left| \rule{0.3cm}{.0cm} \right>}

\def\etal{{\it et al.\ }}

\begin{document}

\title { Theoretical Aspects of Radium-Containing Molecules Amenable to Assembly from Laser-Cooled Atoms for New Physics Searches} 

\vspace*{1cm}

\author{Timo Fleig}
\email{timo.fleig@irsamc.ups-tlse.fr}
\affiliation{Laboratoire de Chimie et Physique Quantiques,
             IRSAMC, Universit{\'e} Paul Sabatier Toulouse III,
             118 Route de Narbonne, 
             F-31062 Toulouse, France }
\author{David DeMille}
\email{ddemille@uchicago.edu}
\affiliation{University of Chicago, Department of Physics, Chicago, IL 60637}
\vspace*{1cm}
\date{\today}

\vspace*{1cm}
\begin{abstract}
We explore the possibilities for a next-generation electron-electric-dipole-moment experiment using ultracold heteronuclear diatomic molecules 
assembled from a combination of radium and another laser-coolable atom.
In particular, we calculate their ground state structure and their sensitivity to parity- and time-reversal (${\cal{P,T}}$) violating physics 
arising from flavor-diagonal charge-parity (${\cal{CP}}$) violation. Among these species, the largest 
${\cal{P,T}}$-violating molecular interaction constants --- associated 
for example with the electron electric dipole moment 
--- are obtained for the combination of radium (Ra) and silver (Ag)
atoms. A mechanism for explaining this finding is proposed. We go on to discuss the prospects for an electron EDM search using ultracold, assembled, optically trapped RaAg molecules, and argue that this system is particularly promising for rapid future progress in the search for new sources of ${\cal{CP}}$ violation.
\end{abstract}

\maketitle
\section{Introduction}
\label{SEC:INTRO}
The detection of a charge-parity- (${\cal{CP}}$) violating signal of leptonic or semi-leptonic origin would open a route \cite{Sakharov_JETP1967,ramsey-musolf_review1_2013} for explaining so far not understood
aspects of the observed matter and energy content of the universe, in particular its matter-antimatter
dissymmetry \cite{Dine_Kusenko_MatAntimat2004}. Under the assumption that ${\cal{CPT}}$ invariance (${\cal{T}}$ denoting time reversal)
of fundamental physical laws holds \cite{pauli_lorentz_CPT}, the detection of an
electric dipole moment (EDM) along the angular momentum of any system would reveal the influence of ${\cal{CP}}$-violating interactions.  EDMs are very insensitive to the ${\cal{CP}}$-odd phases
already incorporated into the Standard Model (SM) of elementary particles (via flavor mixing matrices), so EDMs act as very low background signals for beyond SM ${\cal{CP}}$-odd interactions \cite{Pospelov:1991zt,Yamanaka_leptonEDM_2021}. 
For this reason, atomic and molecular searches for flavor-diagonal violations of 
$\cal{CP}$ symmetry
\cite{FlavorPhysLepDipMom_EPJC2008} have become a field of intense research at the forefront
of New Physics (NP) searches 
\cite{ACME_ThO_eEDM_science2014,Heckel_Hg_PRL2016,Cairncross_Ye_NatPhys2019}.  
In this paper, we focus on ${\cal{P,T}}$-violating effects that explicitly couple to electron spin---in particular, the electron EDM, nucleon-electron scalar-pseudoscalar (Ne-SPS) coupling, and nuclear magnetic quadrupole moment (NMQM).  The sensitivity of a given atomic or molecular species to these effects can be parameterized in terms of the associated ${\cal{P,T}}$-violating interaction constants: the effective internal field acting on the electron EDM, $E_{\text{eff}}$; the Ne-SPS interaction constant $W_{S}$; and the NMQM interaction constant $W_{M}$. We refer to the entire set of these interaction constants as ``the ${\cal{P,T}}$-odd constants''.

Among various future directions considered to search for ${\cal{P,T}}$-odd effects coupled to the electron spin with greater sensitivity \cite{safronova_rmp_2018}, experiments based on ultracold and optically trapped molecules 
\cite{CarrDeMille_NJP2009,Bohn_Rey_Ye_Science2017} 
appear particularly promising \cite{Fitch_Tarbutt2020,Kozyryev_Hutlzer2017,Cairncross_Ye_NatPhys2019}. Here, the structure of polar molecules amplifies the observable energy shifts due to underlying mechanisms for ${\cal{CP}}$ violation \cite{sandars_PRL1967, Sushkov_Flambaum_Khriplovich1984}.
Optical trapping could provide long spin coherence times \cite{Doyle_Burchesky_RotationalCoherence_2021,Zwierlein_Park_SpinCoherence_2017,Cornish_Gregory_SpinCoherence_2021} for large molecular ensembles \cite{Moses_Ye_NatPhys2017}, and hence unprecedented energy resolution. With plausible projected values of experimental parameters, this could provide $\sim$3 orders of magnitude improved statistical sensitivity relative to the current state of the art for the ${\cal{P,T}}$-odd constants of interest here \cite{Kozyryev_Hutlzer2017}. There are significant advantages in using molecules at the lowest possible temperatures, i.e. near the regime of quantum degeneracy. Here, possible systematic errors due to the trapping light can be minimized by using weak, low-intensity trapping light \cite{Romalis_Fortson_EDMLattice_1999,Chu_Chin_CsLatticeEDM_2001}. Moreover, the high and deterministic densities typical of lattice-trapped quantum gases \cite{Bloch_Gross_Lattices_2017} open the potential to employ spin-squeezing methods to surpass the standard quantum limit of statistical sensitivity \cite{Quantum_Metrology_RevModPhys_2018} (which is already typically reached in EDM experiments, and assumed in the estimate above). In principle, squeezing could enable sensitivity improved by up to another $\sim$1-3 orders of magnitude \cite{Quantum_Metrology_RevModPhys_2018,Kasevich_Hosten_Squeezing_2016,Vuletic_Colombo_Squeezing_2021}.

Implementing this vision requires identifying suitable molecular species---that is, species with large values of the ${\cal{P,T}}$-odd constants, and which also plausibly can be trapped and cooled to near the regime of quantum degeneracy. To date, discussion of potential species with these properties has centered on paramagnetic molecules with structure suitable for direct laser cooling, such as YbF or YbOH or RaF \cite{Fitch_Tarbutt2020,Kozyryev_Hutlzer2017,Isaev_CoolingRaF_2010,Titov_Kudashov_RaFEnhancement_2014}.  However, the coldest and densest molecular gases to date have been produced not by direct laser cooling, but instead by assembly of diatomic species from pairs of ultracold atoms \cite{Ye_DeMarco_DegenerateKRb_2019,Moses_Ye_NatPhys2017}.  An early investigation of the prospects for EDM experiments with such systems was made by Meyer \textit{et al.} \cite{Meyer_Bohn_2009}. They considered the neutral species RbYb and CsYb, which have the unpaired electron needed for high sensitivity to the electron EDM and which could be assembled from atoms routinely cooled to quantum degeneracy.  However, they found that the values of $E_{\text{eff}}$ in these molecules were much smaller than expected from simple scaling arguments---in each case, $E_{\text{eff}} < 1  \frac{\text{GV}}{\text{cm}} $. This is roughly two orders of magnitude smaller than $E_{\text{eff}}$ in ThO, the species used by the ACME experiment to place the best current limit on the electron EDM \cite{ACME_ThO_eEDM_nature2018}.  To our knowledge, the idea to use ``assembled'' ultracold molecules for EDM experiments has since not been discussed further in the literature.

A suitable molecular species to be assembled and used to measure the electron EDM must satisfy several criteria.  Naturally, both constituent atoms must be amenable to laser cooling and trapping. For large values of the ${\cal{P,T}}$-odd constants of interest here, the molecule must have an unpaired electron spin in its absolute ground state.  These two criteria together suggest using molecules where one atom has alkali-like structure (single unpaired electron), and the other has alkaline earth-like structure (closed electron shell).  Because the values of the ${\cal{P,T}}$-odd constants scale roughly as $Z^3$ (where $Z$ is the atomic number) \cite{sandars1966enhancement,Bouchiat_Parity_PartI_1974,Flambaum_Khriplovich1985}, at least one of the atoms should be very heavy to maximize their values. Though less critical, it is experimentally convenient to use species that can be strongly polarized in small electric fields; this can be enabled by a large molecular dipole moment and/or small molecular rotational splitting \cite{Sunaga_Ra-A_2019}.

In 2018, the present authors presented the RaAg molecule \cite{Fleig_Mainz2018} 
as a very promising ultracold molecular system for electron EDM searches.
Use of the alkaline earth atom radium (Ra) as the required heavy nucleus for such a future
experiment is strongly suggested, since Ra ($Z=88$) is the heaviest atom where laser cooling and trapping has been demonstrated \cite{guest2007laser}.
The choice of the silver (Ag) atom rather than a true alkali atom as the bonding partner for Ra is less obvious. However, the coinage metals (Cu, Ag, Au) have a nominally alkali-like structure, with one valence $s$ electron above closed shells, so they are in principle amenable to laser cooling.  Indeed, laser cooling and trapping of Ag atoms was demonstrated already over 20 years ago \cite{Uhlenberg_silver_2000}. In addition, the coinage metals have much larger electron affinities \cite{EA_Ag,EA_Au} than the alkalis. Hence, we anticipated that they might form a strong polar bond with the highly polarizable Ra atom \cite{Lim_Schwerdtfeger_2004,bast_polarizabilities_2008}. This type of bond is generically correlated both with a large effective electric field on the electron EDM \cite{hinds_PS1997,Flambaum_Khriplovich1985}, and with a large molecular dipole moment. Large molecular dipole moments in Ra-coinage metal molecules, discussed here, have also been 
found in Refs. \cite{Sunaga_Ra-A_2019,Tomza_2021}.
Sunaga \textit{et al.} \cite{Sunaga_Ra-A_2019} 
discussed properties of radium-A molecules---where A is a halogen or a coinage-metal atom---relevant to molecular electron EDM searches. A more encompassing view on the possibilities of using ultracold diatomic molecules assembled from laser-coolable atoms, however, 
was not discussed in that paper.  Here, we present a comparative study of the effective electric field $E_{\text{eff}}$ acting on the electron EDM in radium-X molecules, where X is a (potentially) laser-coolable alkali or coinage metal atom.

The following section summarizes the theory underlying the presented results on molecular structure. Section \ref{SEC:RESULTS} contains a comparative study of a systematic series of Ra-alkali and Ra-coinage-metal diatomics with an emphasis on ${\cal{P,T}}$-odd and spectroscopic properties. In the final section we conclude, mention ongoing work \cite{Fleig_DeMille_AgRa},
and lay out some prospects for the very near future.  

\section{Theory}
\label{SEC:THEORY}
\subsection{General Definitions and Wavefunctions}

The electronic many-body states of all of the present molecules are denoted as 
$\left| \Omega \right>$ with $\Omega = |M_J|$. These states are represented by relativistic configuration 
interaction wavefunctions
\begin{equation}
        \left| \Omega \right> \equiv \sum\limits_{I=1}^{\rm{dim}{\cal{F}}^t(M,n)}\,
                                       c_{(\Omega),I}\, ({\cal{S}}{\overline{\cal{T}}})_I \rvac
\end{equation}
where ${\cal{F}}^t(M,n)$ is the symmetry-restricted sector of Fock space with $n$ electrons in $M$ four-spinors,
${\cal{S}} = a^{\dagger}_i a^{\dagger}_j a^{\dagger}_k \ldots$ is a string of spinor creation operators,
${\overline{\cal{T}}} = a^{\dagger}_{\overline l} a^{\dagger}_{\overline m} a^{\dagger}_{\overline n} \ldots$ 
is a string of creation operators of time-reversal transformed spinors. The determinant expansion coefficients 
$c_{(\Omega),I}$ are generally obtained as described in refs. \cite{fleig_gasci,fleig_gasci2}
by diagonalizing the Dirac-Coulomb Hamiltonian, in a.u.
\begin{equation}
 \hat{H}^{\text{Dirac-Coulomb}} 
      = \sum\limits^n_j\, \left[ c\, \boldsymbol{\alpha}_j \cdot {\bf{p}}_j + \beta_j c^2 
    - \frac{Z}{r_j}{1\!\!1}_4 \right]
        + \sum\limits^n_{j,k>j}\, \frac{1}{r_{jk}}{1\!\!1}_4
 \label{EQ:DC_HAMILTONIAN}
\end{equation}
in the basis of the states $({\cal{S}}{\overline{\cal{T}}})_I \rvac$,
where the indices $j,k$ run over electrons, $Z$ is the proton number, and
$\boldsymbol{\alpha},\beta$ are standard Dirac matrices. The specific models used in the present work will
be discussed in subsection \ref{SUBSEC:CORR}. The calculation of 
properties using the resulting CI eigenvectors is technically carried out as documented in refs. 
\cite{knecht_luciparII,knecht_thesis}. Atoms and linear molecules are treated in a finite sub-double
group of $D^*_{\infty h}$ (atoms) or $C^*_{\infty v}$ (heteronuclear diatomic molecules) which gives
rise to a real-valued formalism in either case \cite{DIRAC_JCP}.
Definitions of the various property operators used in the present work
will be given in the following sections.


\subsection{${\cal{P,T}}$-Odd Properties}
The electron EDM interaction constant is evaluated as proposed in stratagem II of Lindroth et al.
\cite{lindroth_EDMtheory1989} as an effective one-electron operator via the squared electronic
momentum operator. In the present work ${\cal{P,T}}$-violating properties are only calculated in
molecules so with zeroth-order states denoted as $\left| \Omega \right>$
\begin{equation}
 E_{\text{eff}} = \frac{2\imath c}{e\hbar} \left< \Omega \right|
 \sum\limits_{j=1}^n\, \gamma^0_j \gamma^5_j\, {\bf{p}}\,^2_j \left| \Omega \right>
 \label{EQ:EEFF}
\end{equation}
with $n$ the number of electrons and $j$ an electron index. The implementation in the many-body
framework is described in greater detail in reference \cite{fleig_nayak_eEDM2013}. 
The EDM effective electric field is related to the electron EDM interaction constant
$W_d = -\frac{1}{\Omega}\, E_{\text{eff}}$.

A measurement on open-shell molecules also tests $C_S$, the fundamental nucleon-electron scalar-pseudoscalar
(Ne-SPS)
coupling constant for a neutral weak current between electrons and nucleons \cite{khriplovich_lamoreaux}.
In the framework of an effective field theory the Ne-SPS interaction
energy \cite{Flambaum_Khriplovich1985} can be written as
\begin{equation}
 \varepsilon_{\text{Ne-SPS}} = W_S\, C_S
 \label{EQ:H_SPS}
\end{equation}
where
\begin{equation}
 W_{\cal{S}} := \frac{\imath}{\Omega}\,
        \frac{G_F}{\sqrt{2}}\, A\, \left< \Omega \right| \sum\limits^n_{j=1}\, 
        {\gamma^0_j\gamma^5_j\, \rho({\bf{r}}_j)} \left| \Omega \right>
 \label{EQ:W_S}
\end{equation}
is the Ne-SPS interaction constant for the nucleus with $A$ nucleons, $G_F$ is the Fermi constant,
$\gamma$ is an electronic Dirac matrix, and $\rho({\bf{r}})$ is the nuclear density at
position ${\bf{r}}$. The implementation in the molecular framework is documented in
reference \cite{ThF+_NJP_2015}.

The nuclear magnetic quadrupole moment (MQM) interaction constant has been implemented in reference 
\cite{fleig:PRA2016} and can be written as
\begin{equation}
 W_M = \frac{3}{2\Omega}\, 
 \left< \Omega \right| 
             -\frac{1}{3}\,   \sum\limits_{j=1}^n\,
                          \left\{ \left[ \alpha_1(j) \frac{\partial}{\partial r_2(j)} -
                                      \alpha_2(j) \frac{\partial}{\partial r_1(j)}
                                      \right]\, \frac{r_3(j)}{r^3(j)} \right\}
                                      \left| \Omega \right>.
 \label{EQ:CIEXPECVAL}
\end{equation}
In this case, $r_k(j)$ denotes the $k$-th cartesian component of vector ${\bf{r}}$ for particle $j$ ({\it{idem}} for
the Dirac matrices $\boldsymbol{\alpha}$).

\subsection{Other Properties}

The rotational constant is defined for a classical rigid rotor as $B = \frac{\hbar^2}{2I}$ with $I = \mu R$
the moment of inertia in terms of the reduced mass $\mu$ and the internuclear distance $R$.
Thus, in units of inverse length,
\begin{equation}
 B_e = \frac{B}{hc} = \frac{\hbar}{4\pi c\mu R_e^2}.
 \label{EQ:ROTCON}
\end{equation}
$R_e$ is in the present obtained from quantum-mechanical calculations.
\section{Results}
\label{SEC:RESULTS}

\subsection{Computational Details}

\subsubsection{Basis Sets and Molecular Spinors}

Uncontracted Gaussian atomic basis sets have been used for all considered systems: For Ra Dyall's triple-$\zeta$ set 
\cite{dyall_s-basis,5fbasis-dyall-ccorr} including outer-core correlating functions, amounting to \{33s,29p,18d,12f,3g,1h\};
For Li and Na the EMSL aug-cc-pVTZ sets \cite{EMSL-basis2019};
For K, Rb, Cs and Fr  Dyall's TZ bases including $(n-1)s,(n-1)p,ns$-correlating functions 
(also $(n-2)d$-correlating for Cs and Fr) \cite{dyall_s-basis}.

Molecular spinors are obtained from Dirac-Coulomb Hartree-Fock (DCHF) calculations using the 
\verb+DIRAC+ program package \cite{DIRAC_JCP} in a locally modified version.
Since the present systems have an odd total number of electrons
fractional occupation is used for defining the Fock operator: $f=0.5$ per spinor for one Kramers pair
and the open-shell spinor pair is a molecular superposition of radium-alkali (RaA) atomic contributions.
For Ra-coinage-metal (RaC) molecules the fractional occupation is $f=0.75$ per spinor for the two Kramers pairs denoted 
$\sigma$ and $\sigma^*$ in Fig. \ref{FIG:COINAGE-RA}.

\subsubsection{Correlated Wave Functions}
\label{SUBSEC:CORR}

The Generalized Active Space (GAS) technique \cite{fleig_gasmcscf,olsen_cc} is ensuingly exploited for efficiently 
taking into account leading interelectron correlation effects. The model space generally includes all spinors 
required to describe the molecular ground state including leading electron correlation effects.

\begin{figure}[t]
	\caption{Wavefunction definitions for alkali-radium diatomic molecules. Up to two holes are allowed in
	the sub-valence spinors which accounts for correlation effects among the sub-valence electrons and with
	the valence electrons. The model space is restricted to the valence spinors where all occupations are
	allowed. The cutoff for the virtual space is set to $10$ [a.u.]}
 \includegraphics[angle=0,width=10.0cm]{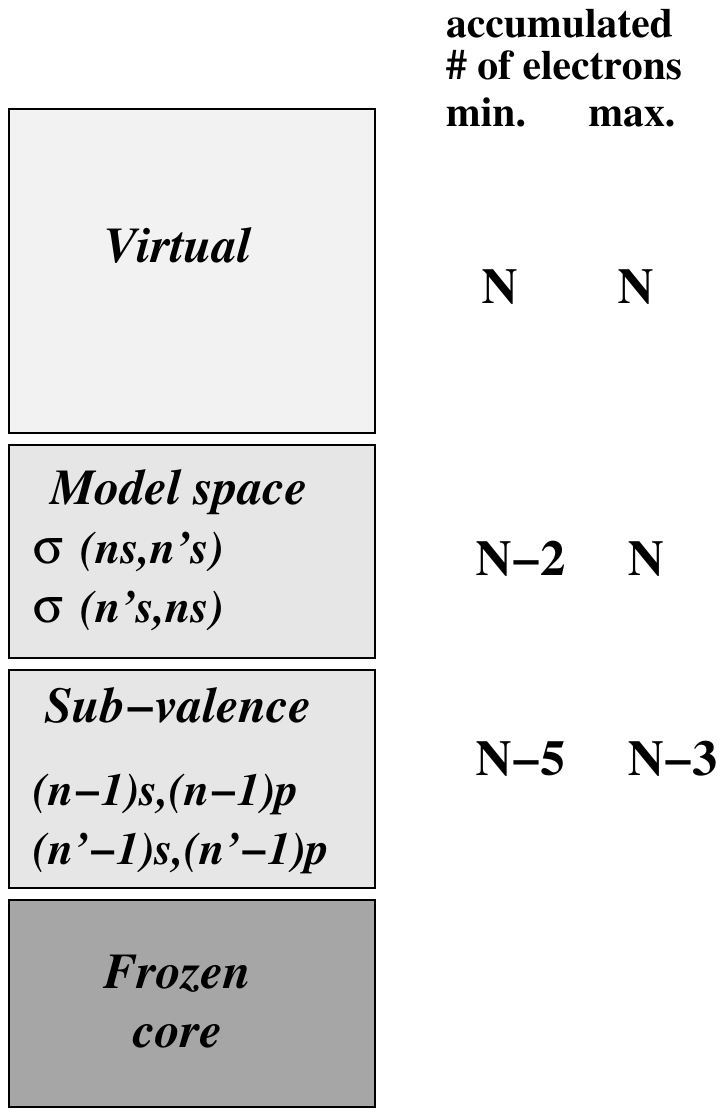}
 \label{FIG:ALKALI-RA}
\end{figure}

For the RaA calculations a specific model is adopted that allows for an economic description of the molecular 
ground state including electron-correlation effects. Fig. \ref{FIG:ALKALI-RA} shows how the wavefunction is linearly
parameterized for this set of calculations.

\begin{figure}[t]
        \caption{Wavefunction definitions for coinage-metal-radium diatomic molecules. 
	Up to one hole is allowed in
        the outer-core spinors which accounts for core-valence correlations between those electrons and the
	respective valence electrons. In the model space all occupations within the given constraints are allowed. 
	$\sigma$ is a molecular spinor pair with predominantly Cu 4s / Ag 5s / Au 6s and some Ra 7s character. 
	$\sigma^*$ is a molecular spinor pair with predominantly Ra 7s and some Cu 4s / Ag 5s / Au 6s character. 
	The cutoff for the virtual space is set to $5$ [a.u.] (CuRa) and $4$ [a.u.] (RaAg and RaAu).}
 \includegraphics[angle=0,width=10.0cm]{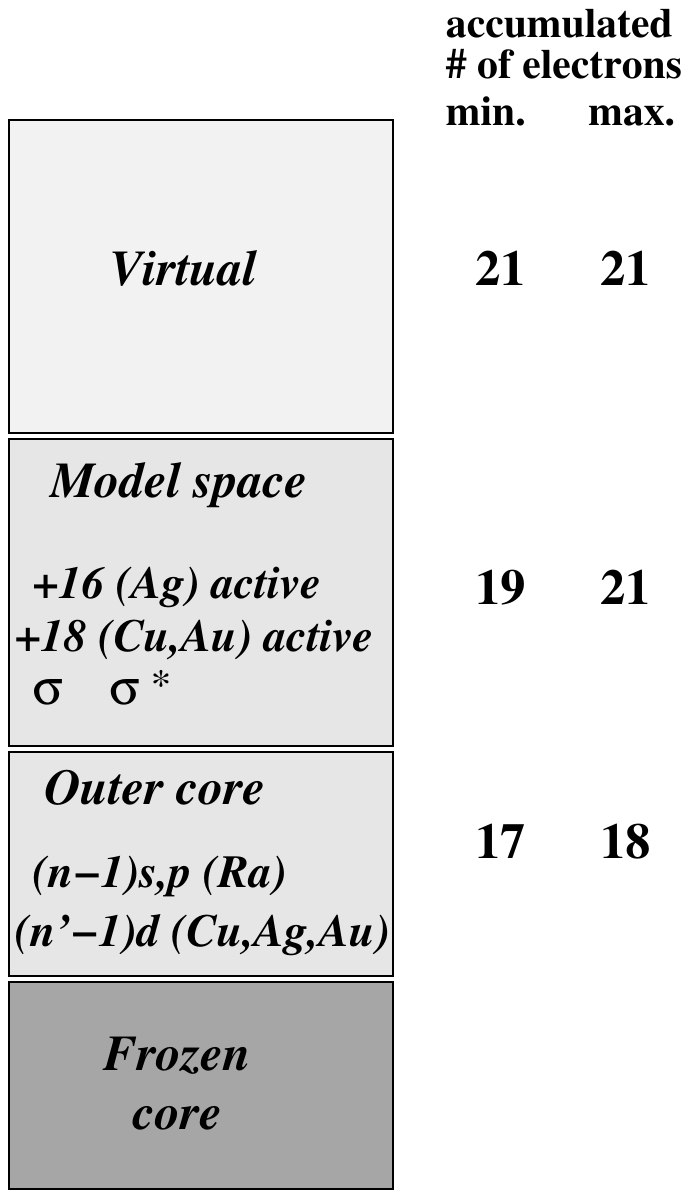}
 \label{FIG:COINAGE-RA}
\end{figure}

For the RaC calculations a different model has been chosen which is shown in Fig. \ref{FIG:COINAGE-RA}.

\subsubsection{Rovibrational properties}

Reduced masses are obtained for the most abundant isotopes $^7$Li, $^{23}$Na, $^{39}$K, $^{85}$Rb, $^{133}$Cs,
$^{223}$Fr, $^{107}$Ag, and $^{226}$Ra with data taken from ref. \cite{AME_2020_data}.

\subsection{Alkali- and Coinage-Metal-Radium Molecules}

\subsubsection{Trends}

The purpose of this section is to show trends for properties of interest among diatomic molecules that are candidates for
measurement of a ${\cal{P,T}}$-odd signal in the lepton sector. The interaction constants for their main sensitivity in this
regard are compared for alkali (A) atoms and coinage-metal (C) atoms bound to atomic radium, see Table \ref{TAB:CONSTANTS}.
Among the A-Ra molecules the
trends are consistent for $E_{\text{eff}}$, $W_{S}$, and $W_{M}$ where in all cases the interaction constants increase
monotonically with decreasing nuclear charge of A. These constants have been obtained at equilibrium internuclear separation
$R_e$, along with other spectroscopic properties and the molecule-frame electric dipole moment $D$. It is worth noting that the root mean-square radii for the $ns$ electron of the atoms A or C, here obtained from atomic quantum-mechanical calculations, are a good predictor of $R_e$, in the sense that the ratio $\frac{R_e \text{[a.u.]}}{
\sqrt{\left< {\hat{r}}^2 \right>_{n{\text{s}}}} \text{[a.u.]}}$ is nearly constant across all the considered molecules. The greatest deviations from the mean value of this ratio amount to around $7$\% for Na (upper end) and Ag (lower end).

Since analytical relationships between the matrix elements of ${\cal{P,T}}$-odd interactions exist 
\cite{PhysRevA.85.029901} and these relationships have also been corroborated in numerical studies of various complex 
systems \cite{Fleig_Jung_Xe_2021,Fleig_Skripnikov2020} the trends for these interactions are expected to be very 
similar which is confirmed by the results in Table \ref{TAB:CONSTANTS}.  The principal
\begin{table}[h]
 \caption{\label{TAB:CONSTANTS} Equilibrium internuclear distances $R_e$, harmonic vibrational frequencies $\omega_e$, rotational constants
	$B_e$, molecule-frame static electric dipole moment $D$, polarizing external field $E_{\text{pol}} = \frac{2 B_e}{D}$, electron EDM effective
	electric field $E_{\text{eff}}$, nucleon-electron scalar-pseudoscalar interaction constants $W_S$, and nuclear magnetic-quadrupole
	moment interactions constants $W_M$ for the electronic ground states {$^2\Sigma_{1/2}$} of diatomics RaA and RaC; For RaLi two sets of results are shown using two different cutoff energies for the virtual spinor set; Experimental electron
	affinities (EA) and root mean-square radius for the valence $s$ electron spinor $\sqrt{\left< {\hat{r}}^2 \right>_{n{\text{s}}}}$ in a.u. for alkali and coinage-metal atoms.
	${\cal{P,T}}$-odd constants are evaluated at the respective $R_e$.
}

 \vspace*{0.4cm}
 \hspace*{-2.4cm}
 \begin{tabular}{l|cccc|c|c|ccc|c}
	 &  $R_e$ [a.u.]  &  $\omega_e$ [\cm]  &  $B_e$ [\cm] &  $D$[Debye]  &  $\sqrt{\left< {\hat{r}}^2 \right>_{n{\text{s}}}}$ &  EA [eV]  &  $E_{\text{eff}} \left[\frac{\rm GV}{\rm cm}\right]$  & $W_{S}$ [kHz]  & $W_{M}$ [$\frac{10^{33} {\text{Hz}}}{e\, {\text{cm}}^2}$] & $E_{\text{pol}} \left[\frac{\rm kV}{\rm cm}\right]$   \\ \hline
 RaLi(10au)  &   $7.668$      &   $105.4$   &   $0.151$ & $1.36$ & $4.21$  & $0.618$ \cite{EA_Li}            &  $22.2$   &   $-59.5$   &   $0.652$ & $13.3$  \\
 RaLi(50au)  &   $7.689$      &   $103.8$   &   $0.150$ & $1.34$ & $4.21$  & $0.618$ \cite{EA_Li}            &  $21.7$   &   $-58.3$   &   $0.641$ & $13.3$  \\
 RaNa        &   $8.703$      &   $ 39.3$   &   $0.038$ & $0.51$ & $4.54$  & $0.548$ \cite{na_exp_eAff_1985} &  $12.0$   &   $-32.2$   &   $0.368$ & $8.90$  \\
 RaK         &   $10.37$      &   $ 20.7$   &   $0.017$ & $0.39$ & $5.60$  & $0.501$ \cite{EA_K}             &  $5.44$   &   $-14.6$   &   $0.167$ & $5.18$  \\
 RaRb        &   $10.75$      &   $ 14.5$   &   $0.008$ & $0.36$ & $5.93$  & $0.486$ \cite{EA_Rb}            &  $5.01$   &   $-13.6$   &   $0.152$ & $2.75$  \\
 RaCs        &   $11.25$      &   $ 12.0$   &   $0.006$ & $0.46$ & $6.48$  & $0.472$ \cite{EA_Cs}            &  $4.52$   &   $-12.6$   &   $0.138$ & $1.48$  \\
 RaFr        &   $11.26$      &   $ 10.5$   &   $0.004$ & $0.24$ & $6.31$  & $0.486$ \cite{landau_EA_Alkali} &  $3.44$   &   $-12.4$   &   $0.137$ & $2.06$  \\ \hline
 RaCu                           &   $6.050$      &   $106.7$   &   $0.033$ & $4.30$  & $3.54$  & $1.236$ \cite{EA_Ag}            &  $67.0$   &   $-180.6$  &   $1.771$ & $0.92$  \\
 RaAg \cite{Fleig_Mainz2018} &   $6.241$      &   $ 90.0$   &   $0.021$ & $4.76$  & $3.73$  & $1.304$ \cite{EA_Ag}            &  $63.9$   &   $-175.1$  &   $1.761$ & $0.53$  \\
 RaAu                           &   $5.836$      &   $98.4$    &   $0.017$ & $5.71$  & $3.30$  & $2.309$ \cite{EA_Au}            &  $50.4$   &   $-166.4$  &   $1.752$ & $0.36$  \\
 \end{tabular}
\end{table}
mechanism explaining this trend becomes obvious when considering the electron affinities (EA)
of A. A free Ra atom has no unpaired electrons and a {$^1S_0 (7s^2)$} ground state insensitive to the present 
${\cal{P,T}}$-odd interactions. In the molecular environment, however, the partner atom A will draw electron density
from Ra leading to effective spin density on the latter. This effect is a function of EA(A) and manifests itself in
non-zero ${\cal{P,T}}$-odd interactions.

\begin{figure}[t]
 \begin{center}
  \includegraphics[angle=0,width=18.0cm]{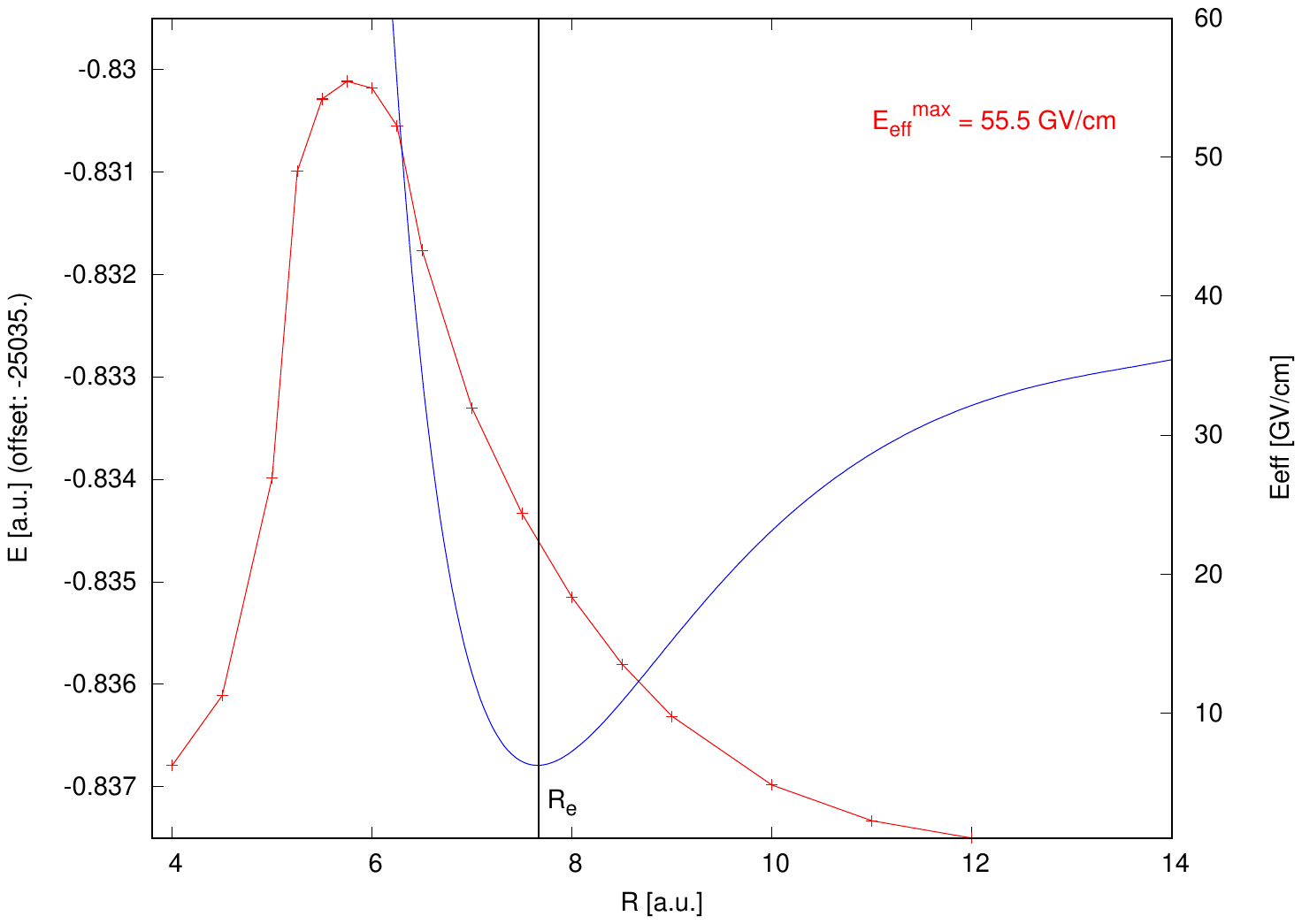}
	 \caption{\label{FIG:EEFF_LIRA} $X^2\Sigma_{1/2}$ potential-energy curve (blue) and $E_{\text{eff}}$ (red curve) against
	 internuclear separation for RaLi. }
 \end{center}
\end{figure}

\begin{figure}[t]
 \begin{center}
  \includegraphics[angle=0,width=18.0cm]{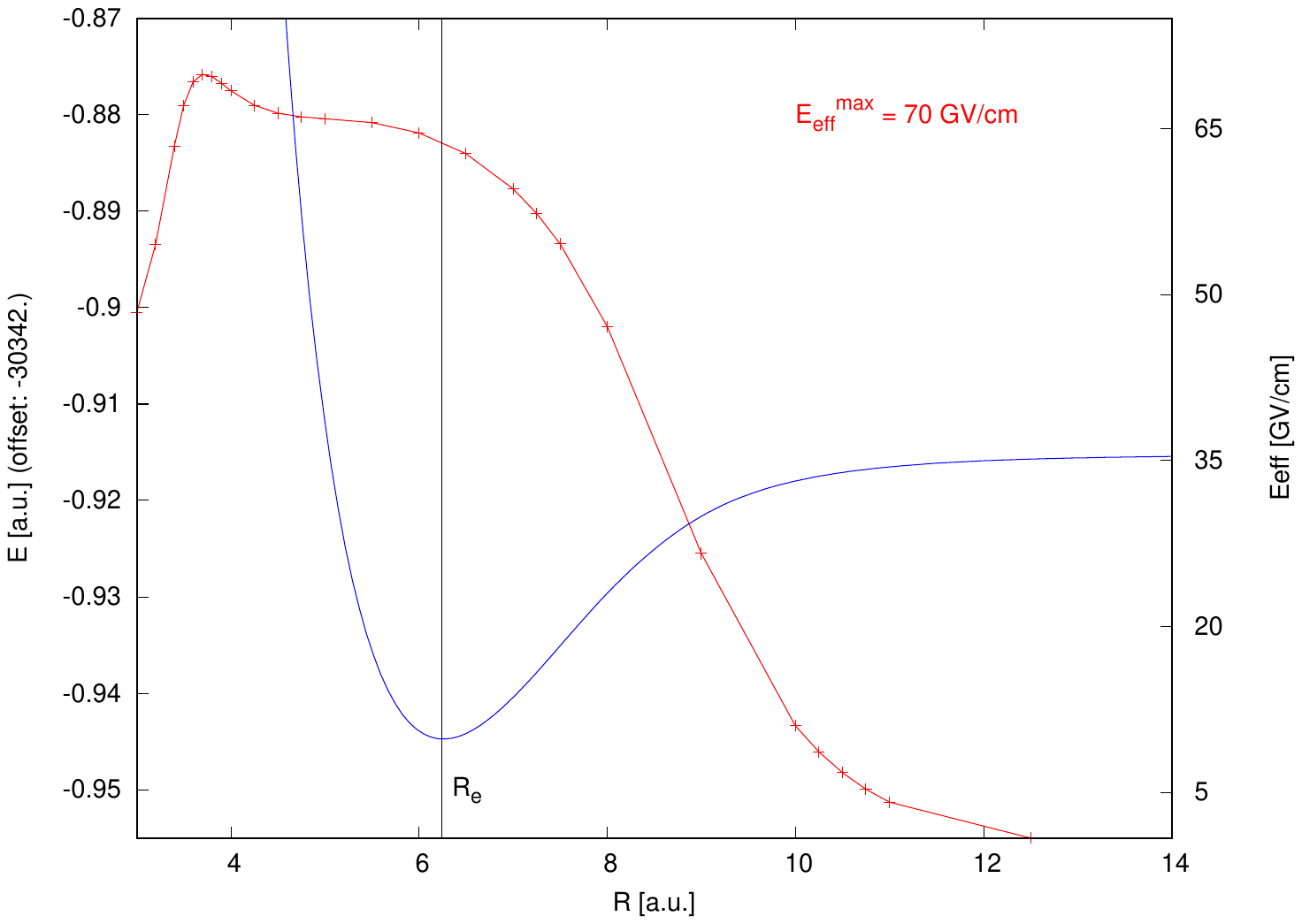}
	 \caption{\label{FIG:EEFF_AGRA} $X^2\Sigma_{1/2}$ potential-energy curve (blue) and $E_{\text{eff}}$ (red curve) against
	 internuclear separation for RaAg}
 \end{center}
\end{figure}

This mechanism of creating spin density on Ra is qualitatively the same for all Ra-A combinations. 
However, there exist pronounced quantitative differences for the different partner atoms A, leading
to sizeable differences in $E_{\text{eff}}$ at the equilibrium internuclear separation of the respective
molecule. Figure \ref{FIG:EEFF_LIRA} shows that $E_{\text{eff}}$ goes through a maximum at separations
shorter than $R_e$ for RaLi and drops off quite sharply as the molecule is stretched beyond $R_e$.
The corresponding situation for RaAg is displayed in Fig. \ref{FIG:EEFF_AGRA}. In contrast to RaLi
the RaAg curve for $E_{\text{eff}}$ hardly drops off from the maximum value as $R$ passes through
the minimum of the potential-energy curve (PEC) but instead displays a shoulder that extends to values $R > R_e$. Even though
${\text{E}}_{\text{eff max}}$(RaAg) $\approx 70 \frac{\rm GV}{\rm cm}$ is only about
$25$\% greater than ${\text{E}}_{\text{eff max}}$(RaLi) $\approx 56 \frac{\rm GV}{\rm cm}$,
the shoulder leads to almost a factor of $3$ difference between $E_{\text{eff}}$(RaAg) and $E_{\text{eff}}$(RaLi) at the respective values of $R_e$.

\begin{figure}[t]
 \begin{center}
  \includegraphics[angle=0,width=18.0cm]{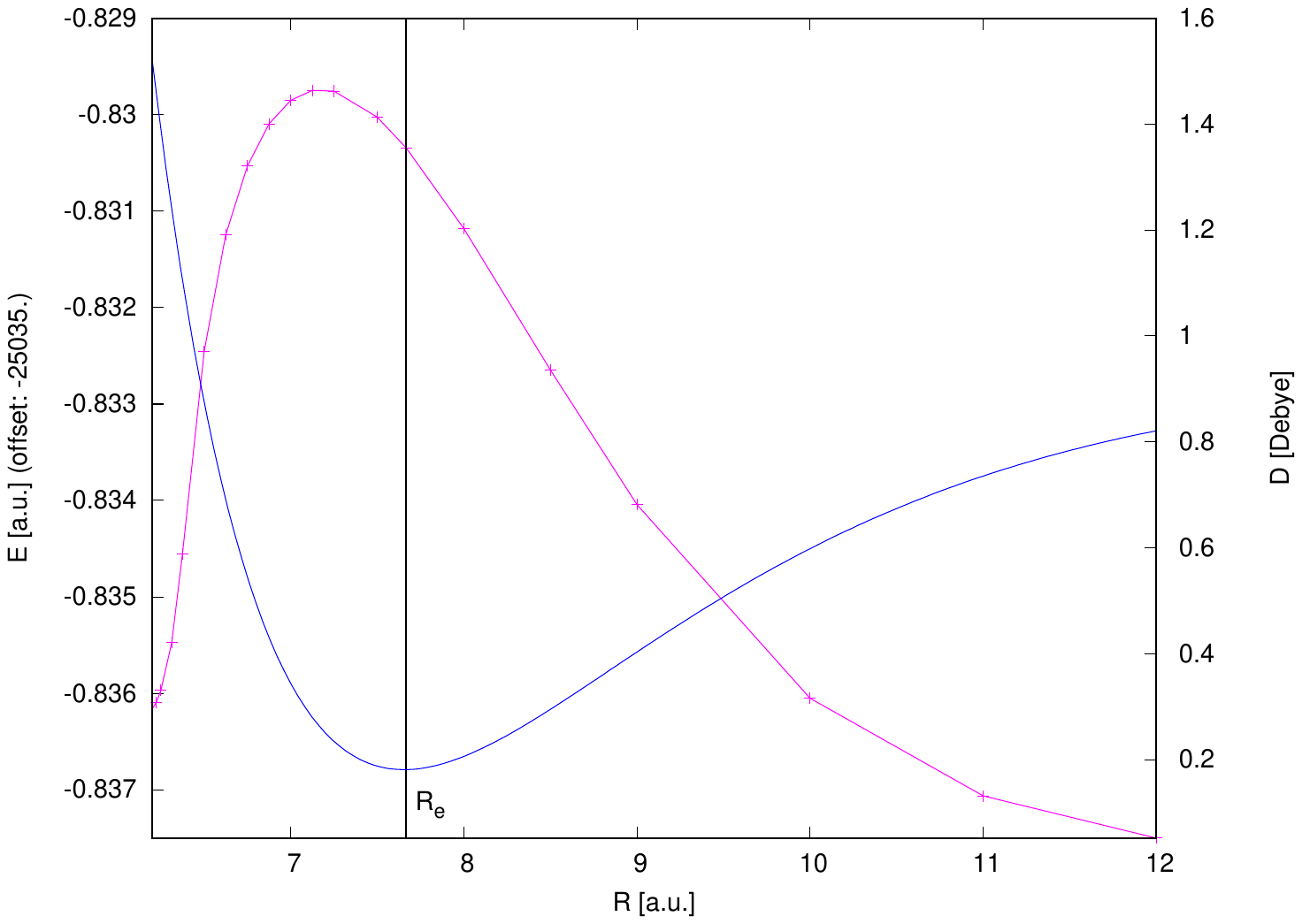}
         \caption{\label{FIG:D_LIRA} $X^2\Sigma_{1/2}$ potential-energy curve (blue) and molecule-frame electric
	 dipole moment $D$ (magenta curve) against internuclear separation for RaLi (cutoff 10 a.u.)}
 \end{center}
\end{figure}

\begin{figure}[t]
 \begin{center}
  \includegraphics[angle=0,width=18.0cm]{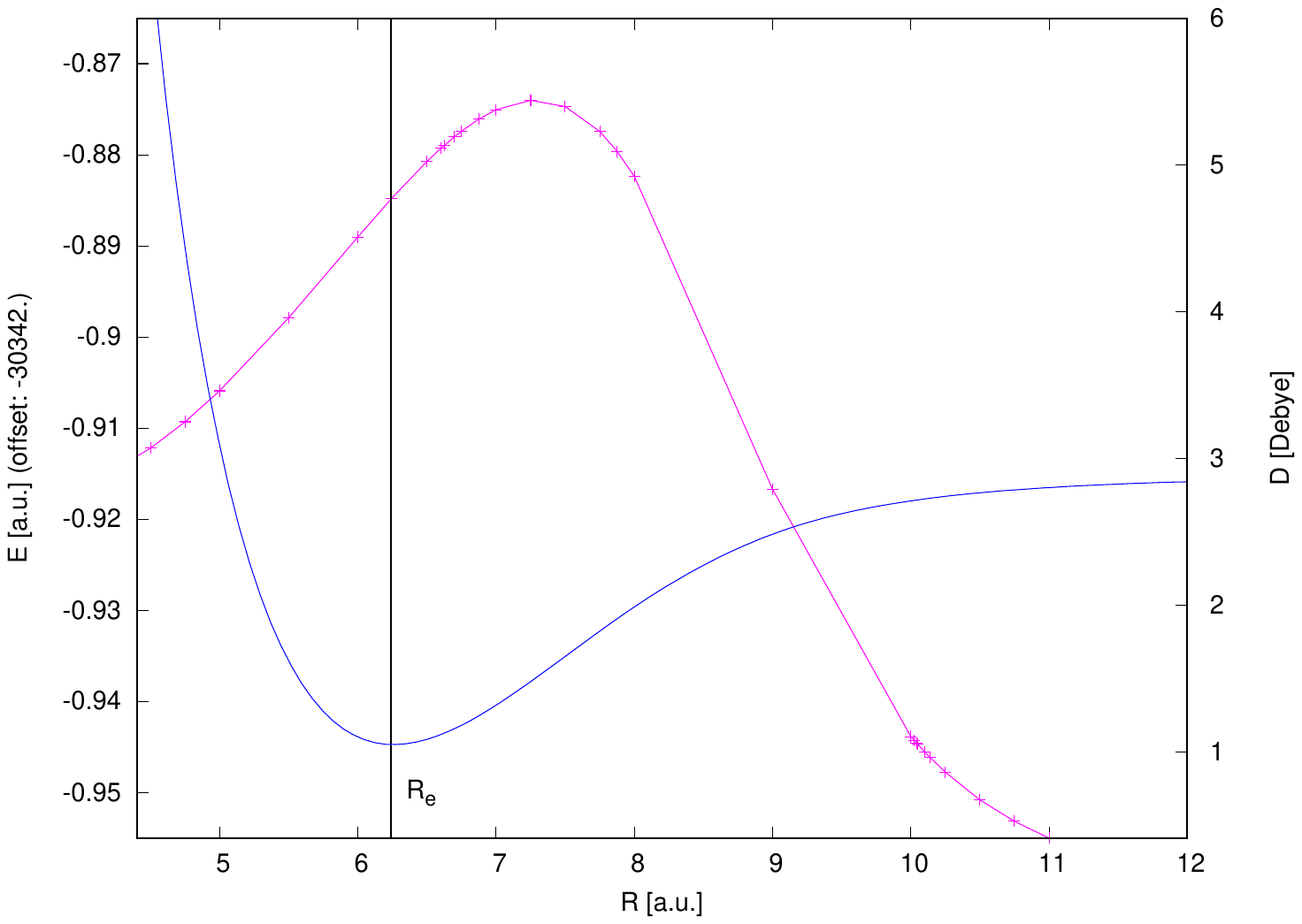}
         \caption{\label{FIG:D_AGRA} $X^2\Sigma_{1/2}$ potential-energy curve (blue) and molecule-frame electric
	 dipole moment $D$ (magenta curve) against internuclear separation for RaAg}
 \end{center}
\end{figure}

With the present electronic-structure models we find 
$R_{{\text{E}}_{\text{eff max}}} = 5.75$ a.u. and $R_e = 7.69$ a.u. for RaLi. 
The change of molecule-frame
EDM between these two points is $\Delta D = D(R_e) - D(R_{{\text{E}}_{\text{eff max}}}) = 0.84$ Debye,
and the full dipole-moment curve is shown in Fig. \ref{FIG:D_LIRA}. 
This is a positive but rather modest value. On the contrary, for RaAg $R_{{\text{E}}_{\text{eff max}}} =
3.7$ a.u., $R_e = 6.24$ a.u. and $\Delta D = D(R_e) - D(R_{{\text{E}}_{\text{eff max}}}) = 1.8$ Debye,
as shown in Fig. \ref{FIG:D_AGRA}. 
In addition to the EDM being much greater at $R_e$ in RaAg, it also displays a sharper increase between
the two significant points, indicating that the partial charge remains on the Ag atom even when the
internuclear distance is stretched slightly beyond $R_e$. This leads to a shoulder both for the spin
density on the Ra atom and $E_{\text{eff}}$ in RaAg. The underlying principal explanation is, therefore, strongly
suggested to indeed be the electron affinity of the atom polarizing the heavy target atom (Ra).

A further analysis shows that the partial charge $\delta^-_A$ on the atom A
at $R_e$ calculated from DCHF valence $s$ spinors increases (on the absolute) from 
$\delta^-_{\text{Fr}} \approx -0.03\, e$ to $\delta^-_{\text{Rb}} \approx -0.05\, e$ to 
$\delta^-_{\text{Li}} \approx -0.08\, e$ and reaches $\delta^-_{\text{Ag}} \approx -0.24\, e$ in RaAg, leading
to a significantly greater $E_{\text{eff}}$ in RaAg than in RaLi. 

Turning to a comparison of the coinage-metal-Ra molecules it becomes clear that the above analysis in terms
of EA alone is not sufficient for explaining all trends. Au has a much greater EA than Cu or Ag but still
yields a smaller $E_{\text{eff}}$ when bound to Ra. The maximum value for $E_{\text{eff}}$ as a function of
$R$ is greatest for CuRa which also displays the characteristic shoulder as discussed above for the case of
RaAg (although it is less pronounced in RaCu). In the case of RaAu, the shoulder is significantly
less pronounced than in RaAg in addition to the maximum $E_{\text{eff}} (R)$ being smallest among all
coinage-metal-Ra molecules. An explanation for this observation in terms of $s$-$p$-mixing matrix elements and 
relevant spinor energies has been attempted in Ref. \cite{Sunaga_Ra-A_2019}. We refrain from delving into a
deeper analysis of the related trends for $E_{\text{eff}} (R)$ in coinage-metal-Ra molecules since, as explained below, neither
Cu nor Au atoms have significant advantages, relative to Ag, for use in an ultracold assembled molecule EDM experiment.

We also consider 
a further important aspect
for experimental feasibility: the external electric field $E_{\text{pol}} = 2B_e/D$ required for fully polarizing the molecule. The comparison in Table \ref{TAB:CONSTANTS} demonstrates that RaAg is among the best of the laser-coolable atom combinations also in this respect (surpassed only by RaAu, by a factor of only 1.5). The extremely small $E_{\text{pol}}$ required
for RaAg is due to its much greater molecule-frame EDM -- about an order of magnitude -- as compared with 
heavier alkali-radium molecules. This by
far outweighs the slight disadvantage RaAg has in terms of its rotational constant $B_e$, which is roughly
a factor of three greater than $B_e$ for RaCs, as an example. A similar conclusion was reached in Refs.~\cite{Sunaga_Ra-A_2019,Tomza_2021}.

\subsubsection{Comparison of Models and with the Literature}

The studies in the previous section have a focus on establishing trends and explaining qualitatively important
features of the candidate set of molecules. For RaAg the corresponding wavefunction model is defined in
Fig. \ref{FIG:COINAGE-RA} and we will here call it TZ/MR-CISD. Models of similar quality have been used for the 
other molecules in this work for the mentioned purposes.

In the following we compare the TZ/MR-CISD to a more accurate model QZ/MR-CISDT which differs from the
previous one in two ways: QZ/MR-CISDT uses Gaussian basis sets of quadruple-zeta (QZ) quality 
\cite{dyall_4d,dyall_s-basis} for both Ra and Ag atoms. Second, the model space now only consists of
the Kramers pairs $\sigma$ and $\sigma^*$, but in turn the highest particle rank of the virtual spinor
space has been increased from $2$ to $3$. The model QZ/MR-CISDT is, therefore, significantly more
accurate than the model TZ/MR-CISD for a calculation of the electronic ground state.

\begin{table}[h]
 \caption{\label{TAB:CONSTANTS_COMP} Spectroscopic and ${\cal{P,T}}$-odd constants for RaAg}

 \vspace*{0.4cm}
 \hspace*{-2.4cm}
 \begin{tabular}{l|cccc|ccc|c}
  Source   &  $R_e$ [a.u.]  &  $\omega_e$ [\cm]  &  $B_e$ [\cm] &  $D$[Debye]  &  $E_{\text{eff}} \left[\frac{\rm GV}{\rm cm}\right]$  & $W_{S}$ [kHz]  & $W_{M}$ [$\frac{10^{33} {\text{Hz}}}{e\, {\text{cm}}^2}$] & $E_{\text{pol}} \left[\frac{\rm kV}{\rm cm}\right]$   \\ \hline
 present TZ/MR-CISD  \cite{Fleig_Mainz2018} &  $6.241$  &  $ 90.0$  &  $0.0213$ & $4.76$  &  $63.9$  &  $-175.1$  &  $1.761$ & $0.53$  \\
 present QZ/MR-CISDT &  $6.128$  &  $98.2$  &  $0.0221$ & $4.89$  & $66.1$  &  $-181.1$  &  $1.821$ & $0.54$  \\ \hline
 \'Smia{\l}kowski {\etal} \cite{Tomza_2021} CCSD(T)  &  $5.959$  &  $100.6$  &  $0.0234$ & $5.08$ &  &  &  & $0.55$  \\
 Sunaga {\etal} \cite{Sunaga_Ra-A_2019} CCSD  &   $6.10$  &   &  $0.022$ & $5.1$ &  $73.7$  &  $-201.8$   &   & $0.52$  \\ \hline
 \end{tabular}
\end{table}

Comparative results are shown in Table \ref{TAB:CONSTANTS_COMP}. The results for our spectroscopic and 
${\cal{P,T}}$-odd constants for RaAg from the two presented models differ by less than $5$\%, except for
$\omega_e$ where the difference is around $9$\%. The results from our more accurate model, QZ/MR-CISDT,
differ from those of \'Smia{\l}kowski {\etal} by less than $4$\%.
The results presented in Ref. \cite{Tomza_2021} have been obtained with large valence atomic basis sets and a 
high-quality wavefunction model and are certainly the more reliable values for comparison than those obtained
by Sunaga {\etal} \cite{Sunaga_Ra-A_2019} where basis sets of only double-zeta (DZ) quality have been used.
However, \'Smia{\l}kowski {\etal} use effective core potentials (ECP) whereas our calculations employ
Dirac wavefunctions for the entire set of atomic shells.

Concerning the difference of $E_{\text{eff}}$ with the result from Ref. \cite{Sunaga_Ra-A_2019} we tested a 
basis set of double-zeta quality \cite{dyall_4d,dyall_s-basis} and obtain 
$E_{\text{eff}} (R = 6.0\, {\text{a.u.}}) \approx 72\, \frac{\rm GV}{\rm cm}$
at an internuclear separation close to $R_e$ determined in the work of Sunaga {\etal} which is very close to
the result from Ref. \cite{Sunaga_Ra-A_2019} in Table \ref{TAB:CONSTANTS_COMP}. Since the two employed
wavefunction models (MR-CISDT and CCSD) are similar in quality this demonstrates
that the use of too small a basis set will lead to a significant overestimation of $E_{\text{eff}}$ for RaAg.
$W_{S}$ in Ref. \cite{Sunaga_Ra-A_2019} is, therefore, also too large on the absolute.
It is clear, however, that Ref. \cite{Sunaga_Ra-A_2019} did not aim at highly accurate results but rather
at determining trends for a set of molecules.

\clearpage
\section{Conclusions and Outlook}
\label{SEC:CONCLUSIONS}
In the present work we have shown that radium-coinage metal molecules have much greater ${\cal{P,T}}$-odd interaction constants than Ra-alkali molecules, which are the natural species to consider for assembly from ultracold atoms and high sensitivity to the electron EDM. Moreover, a simple explanation was
developed for how these effects reach near-optimal values in Ra-coinage metal molecules.  We also showed that these RaC species have large intrinsic molecule-frame dipole moments, which make them easily polarized using an external electric field of very modest strength. 

From this perspective alone, any of the Ra-coinage metal molecules could be an interesting experimental system for future EDM experiments.  However, let us return to the original goal, which was to find suitable species that can be assembled from ultracold atoms.  Among the coinage metals (CMs), Ag turns out to be uniquely easy to laser cool and trap. In all the CM atoms, the last filled electron shell contains $d$ orbitals, and the energy to excite one electron from the closed $d$-shell to the unfilled $s$ orbital is quite comparable to that needed to excite the valence electron from its $ns$ state to an $np$ orbital (as desired for laser cooling). 
In both Cu and Au, the lowest $d$-shell excited states lie at least $\sim\!2$~eV below the $np$ valence excited state, and the $np$ state decays with significant branching ratio into these metastable levels \cite{nist_asd_doi}. Hence, laser cooling of Cu and Au would require additional laser(s) to repump the lower states, in addition to the primary laser driving the $ns-np$ optical cycling transition \cite{Dzuba_Schiller2021}. This in turn would create a ``type-II'' level structure, where the unavoidable presence of dark states significantly reduces the strength of optical forces \cite{devlin2016three}.  By contrast, in the Ag atom the $d$-shell excited state is less than 0.025~eV below the $p_{3/2}$ valence excited state, and the branching ratio for decay into the metastable state is effectively negligible \cite{Uhlenberg_silver_2000}. Hence, in Ag a single laser is sufficient to produce maximal trapping and cooling forces, in complete analogy to standard alkali atoms \cite{Uhlenberg_silver_2000}.  For this reason, within the CM group only Ag has been cooled and trapped, and (to our knowledge) laser cooling of Au or Cu has not even been attempted.  Since the values of the $\cal{P,T}$-odd interaction constants are no better in RaAu or RaCu than in RaAg, and the electric field needed to polarize RaAg is small enough to be experimentally convenient, we conclude that, among the considered species, RaAg is by far the most favorable for experiments of the considered type.

This leads to further questions about experimental viability of an electron EDM search using ultracold, assembled RaAg molecules.  To date, no ultracold alkaline earth-alkali metal molecules of \textit{any} species have been assembled.  However, experimentally plausible pathways for such assembly have been identified for analogous species such as RbSr \cite{devolder2021laser,RbSr_dulieu_2018,RbSr_magneto_2010}, YbLi \cite{brue2012magnetically}, and YbCs \cite{guttridge2018production}, and considerable experimental progress has been made with each of these species \cite{barbe2018observation,green2020feshbach,guttridge2018two}.  Moreover, these pathways are based on extensive, successful experience with assembly of bi-alkali molecules \cite{Moses_Ye_NatPhys2017}.  For this reason, we consider it very plausible that RaAg molecules can, with sufficient effort, be assembled from ultracold Ra and Ag atoms.

All known and proposed techniques for ultracold molecule assembly rely on a two step, coherent process \cite{CarrDeMille_NJP2009}.  In the first step, atom pairs are transferred to a weakly-bound molecular state using either a Feshbach resonance \cite{kohler2006production} or near-threshold photoassociation \cite{Stellmer2012,ciamei2017efficient}.  The weakly-bound state is then transferred to the rovibronic ground state, using STimulated Raman Adiabatic Passage (STIRAP) \cite{vitanov2017stimulated}. The relevant coupling strengths are determined by transition dipole moments between vibronic states for optical transitions \cite{Moses_Ye_NatPhys2017}, or by the structure of long-range bound states for Feshbach association \cite{feshbach_rev_2006}. To understand and reliably calculate all relevant coupling strengths, it is necessary to construct full potential-energy curves (PECs), including short-range and long-range internuclear parts, for ground and electronically excited molecular states. To address this question, we will in forthcoming work present predictions of the relevant features for RaAg.  This will include dispersion coefficients for Ra and Ag atoms so far not established in the literature and required for the long-range parts of the relevant PECs, as well as calculations of short-range PECs and analysis of relevant vibronic transition dipole moments. 

In summary: we have identified the radium-silver (RaAg) molecule as an exceptionally
interesting system for a next-generation electron electric-dipole-moment experiment using ultracold, trapped molecules assembled from laser-coolable atoms.  Further work is underway to evaluate details of the molecular structure that will determine the feasibility of Ra+Ag assembly with high efficiency. 
 

\bibliographystyle{unsrt}
\newcommand{\Aa}[0]{Aa}

\clearpage

\end{document}